\documentstyle[preprint,prl,aps]{revtex}
\begin{document}

\draft

\title{
An Impurity Driven Phase Transition in the
Antiferromagnetic Spin-1 Chain}

\author{R. A. Hyman}
\address{
School of Physics,
Georgia Institute of Technology,
Atlanta, Georgia 30332}

\author{Kun Yang}
\address{
Department of Electrical Engineering,
Princeton University,
Princeton, New Jersey 08544} 

\date{\today}

\maketitle
\begin{abstract}
Using an asymptotically exact real space renormalization procedure,
we find that the Heisenberg
antiferromagnetic spin-1 chain undergoes an impurity driven 
second order phase
transition from the Haldane phase to the random singlet phase, as the
bond distribution is broadened.  
In the Haldane phase and near the critical point, there is 
a Griffiths region in which the gap is filled 
and the susceptibility diverges in a non-universal
manner. The correlation length critical exponent is
$\nu\approx2.3$.   
\end{abstract}
Recently, there has been tremendous 
interest in the antiferromagnetic (AF) spin-1 chain, 
inspired by
the famous conjecture by Haldane\cite{haldane}   
that
integer-spin chains
behave quite differently from half-odd-integer-spin chains.    
For example, in the absence of disorder,
the spin-1 chain has short-range spin-spin
correlations in the ground state and an excitation 
gap\cite{haldane} whereas the spin-1/2 chain is critical.    
The ground state of the spin-1 chain also has a novel string-topological 
order~\cite{rn}.
Some of these results have been experimentally confirmed~\cite{renard}.

Randomness is always present in real materials. 
Theoretical work has demonstrated that randomness dramatically
affects the physical properties of the AF spin-1/2 chain\cite{mdh,fisher1}
and other random one-dimensional magnetic 
systems\cite{fisher2,westerberg,hyman,kunspaper}.
In this letter we report a systematic theoretical study
on the effects of bond
randomness on the AF spin-1 chain, based on the real space renormalization
group scheme developed by Ma {\em et al.}\cite{mdh} (see also Ref. \cite{bl})
and extended by Fisher\cite{fisher1}.

Our main result is that in the presence of bond randomness, there are two
distinct phases in the AF Heisenberg spin-1 chain, separated by a {\em second
order} critical point. The nature of these two phases are described
below.

The ground state in the
Haldane phase in the absence of randomness
is well described by the valence bond solid (VBS)
state~\cite{affleck}.  In the VBS state each spin-1 is composed of two
symetrized spin-1/2 objects.  The spin-1/2 objects form singlets with
spin-1/2 objects on neighboring sites.  This state resembles the ground state 
of a 
dimerized
spin-1/2 chain.   
The ground state is nondegenerate, there is an excitation gap, as well as
a very stable topological structure.  
Thus we expect the Haldane phase and its topological structure to be
{\em stable} against weak bond randomness~\cite{hyman}. 
On the other hand, when
the randomness is strong and the distribution of bond strength is broad,
the origonal spin-1 objects coupled by strong bonds form inert singlet 
pairs and 
generate 
effective further neighbor AF couplings. 
An asymptotically exact real space renormalization group (RG) 
analysis\cite{mdh,fisher1} shows 
that in this case the system flows toward a 
random singlet (RS) phase\cite{fisher1} with {\em universal} thermodynamic
properties and power law behavior in {\em averaged}
spin-spin correlations.
In order to study the transition from the Haldane phase to the random singlet
(RS) phase,
we have extended this RG scheme so that it may be used in both phases.
We find the transition between these two phases is {\em second order}.
The extended RG scheme becomes asymptotically exact in the
low energy limit at the critical point, as
well as in the RS phase.
Thus we are able to extract {\em exact} information about the critical point.
For examples, we find as the randomness strength approaches the critical
point from the Haldane phase, the average
spin-spin correlation length diverges 
in a power law manner with exponent $\nu=\frac{6}{\sqrt{13} -1} \approx 2.3$.
The string-topological order parameter vanishes with a power law 
exponent $\gamma =2\nu \approx 4.6$.

Consider the Hamiltonian
\begin{equation}
H=\sum_i{J_i \bf{S}_i\cdot\bf{S}_{i+1}},
\end{equation}
where $\bf{S}_i$ are
spin-S operators 
and $J_i$ are random
coupling constants (assumed positive unless otherwise specified).
Let us begin with the case that randomness is strong and the width of the
distribution of $J$ is very broad (on a logarithmic scale). In this case
we can use the decimation
renormalization group
procedure developed by Ma, Dasgupta, and Hu~\cite{mdh} (MDH) for the special
case of $S=1/2$.
We first pick the largest bond in the system,
say $J_2$ between spins 2 and 3. Since this is
such a strong bond, and the width of the bond distribution is broad,
bonds $J_1$ and $J_3$ will likely be much weaker than $J_2$. Thus
to lowest order in $J_1$ and
$J_3$, spins 2 and 3 form a singlet pair and
become unimportant at low energies (on scales much smaller than
$J_2$). The major physical effect of the existence of spins 2 and 3 is
to
generate an induced coupling between neighboring spins 1 and 4.
$\tilde{H}_{1-4}=\tilde{J}_{14}\bf{S_1}\cdot\bf{S_4}$
where $\tilde{J}_{14}=\frac{2}{3}S(S+1)J_1J_3/J_2+O(1/J_2^2)$.
This
formula is correct even if $J_1$ and/or $J_3$ are ferromagnetic (F) as long
as their magnitude is much less than $J_2$.
The generated bond is typically {\em much weaker} than all three original
bonds.
Thus the effect
of this decimation procedure is to get rid of the strongest
bond (and also its two neighbors)
in the system, generate a weaker bond between the spins neighboring the
decimated ones, and lower the overall energy scale. 
Fisher\cite{fisher1} has shown rigorously that as one proceeds with the
RG scheme 
the width of bond distribution broadens and the accuracy of the procedure
improves as the energy scale is lowered. 
It becomes {\em asymptotically exact} in the long-distance, low-energy limit
where the
bond distribution flows toward an energy scale dependent stable fixed 
point distribution. Fisher names the phase characterized by this universal
bond distribution the random singlet (RS) phase. Thus in the
presence of strong enough bond randomness, a spin-$S$ Heisenberg AF chain
will be in the RS phase, irrespective of the size of $S$ and independent
of $S$ being an integer or a half-odd-integer.

The above real space RG procedure may not be accurate when the initial
bond distribution is not very broad. In this case there is a finite 
probability that $J_1$ and/or $J_3$ is of similar strength as $J_2$, in which
case the lowest order perturbation is not sufficient. What one needs to do 
in this case is to pick the segment in the chain in which all spins are 
coupled by strong bonds (while the segment itself is only weakly coupled 
to the rest of the chain), solve the spectrum of the segment, and keep only
the low energy states, and consider their coupling to the rest of the chain.
In the case of a spin-1/2 chain, the ground state for a segment is either a 
singlet (for even segments) or a doublet (for odd segments), which are
separated from higher energy states by a gap.
In the former case the segment will be inert, and merely mediates a weak
effective AF coupling between the two spins neighboring the segment, while 
in the latter case the segment may be modeled by an effective spin-1/2 at
low energy which is coupled to the rest of the chain antiferromagnetically.
Thus the structure of the RG scheme remains the same as the MDH
procedure, even if 
greater care is necessary in the beginning.
It is therefore believed that the RS phase correctly 
describes the long-distance, low-energy physics of the spin-1/2 chain, even
if there is only {\em weak} randomness\cite{fisher1}.

The situation is very different in the case of the spin-1 chain. 
For a finite segment of the spin-1 chain with no disorder, 
there are two effective
{\em half} spins localized near the two edges of the segment, 
and the coupling between them is $g(l)\sim
(-1)^lJa^l$, where $l$ is the length of the segment and
$a<1$~\cite{hagiwara}. Thus the coupling between the
two half spins in the same segment may be either ferromagnetic
or antiferromagnetic, and decays {\em exponentially} with the length 
of the segment. The coupling of an edge spin to the rest of the chain,
however, remains antiferromagnetic.
The low energy states of a segment are the singlet and 
triplet states formed by these two edge spins
(and both of them must be kept); other states will be separated
from them by the Haldane gap of order $J$. 
We expect this scenario to remain correct even
when there is weak randomness in the bond strength in such finite segments.
Thus even though
the spin chain is composed of  
spin-1 objects, the {\em effective} degrees of 
freedom at low energies are actually half spins. This is also true when
there is no randomness.  The   
valence bond solid state is also composed of half
spins~\cite{affleck}.   
We may describe the low energy physics of a random spin-1 chain using an
an effective Hamiltonian in terms of these half spins, 
with the following structure: the chain consists of half spins {\em only};
the even bonds are taken from an
antiferromagnetic bond distribution, and odd bonds are taken from a
distribution containing antiferromagnetic and ferromagnetic bonds.
Physically the even bonds are couplings between edge half spins of
neighboring segments (which are always AF), and odd bonds are the coupling
between edge spin in the same segment. Such a effective description
is particularly accurate for the special case of a spin-1 chain with a bimodal
bond distribution: most of the bonds are of strength $J$, while a small but
finite fraction of bonds have a much smaller strength $J'$. In this case the
system can indeed be viewed as a collection of weakly coupled segments of
uniform chains. The couplings between edge half spins are random because 
the length of the segments are random. We believe this model correctly describes
the long-distance, low-energy physics of a random spin-1 chain.
In particular, the original spin-1 Hamiltonian may be recovered by setting
all odd bonds strongly ferromagnetic\cite{hida}.

To study this model, we develop an extended version of
the MDH real space renormalization scheme, which properly 
accounts for  
strong ferromagnetic bonds.   
In our scheme, at any stage of RG, the energy scale $\Omega$ is set by
the strongest {\em antiferromagnetic} bond in the
system. All even bonds, which are all AF, will be weaker than 
$\Omega$. We separate the odd bonds, which can be either F or AF, into two
groups: group A consists of all AF bonds and those F bonds that are weaker 
than $\Omega$, 
while group B consists of F bonds that are stronger than $\Omega$.   
The new spin decimation procedure works in the following way. Find
the strongest AF bond in the system, say $J_i$. If $i$ is odd, then its 
neighbors are both {\em weaker} AF bonds. We just follow the usual
MDH decimation procedure. If $i$ is even 
and both neighbors belong to group A 
we can again follow the MDH decimation procedure. If one of the bonds,
say $J_{i+1}$, belongs to group B, we solve the 3-spin
cluster problem 
of $S_i$, $S_{i+1}$ and $S_{i+2}$, and keep the low energy states, which are a 
doublet.
The doublet may be modeled by a new spin-1/2, which couples to the rest of the 
chain.
If both neighbors of $J_i$ belong to group B, we solve the 4-spin 
cluster including spin $i-1$, $i$, $i+1$ and $i+2$. The ground state is a
singlet, with an excitation gap of order $J_i$.   
Thus it drops out at low energy and mediates an effective AF coupling between
spins $i-2$ and $i+3$. It is easy to show that this procedure keeps the 
original structure of the system; i.e., even bonds AF and odd bonds
F or AF.

The flow equations for the distributions of bonds in this
approximate RG scheme are
\begin{eqnarray}
\nonumber
\lefteqn{-{dP_e(J,\Omega)\over d\Omega}=
\left [P_o(\Omega,\Omega)+N^2(\Omega)P_e(\Omega,\Omega)\right]}\\
\nonumber &\cdot& \int_0^\Omega dJ_1P_e(J_1,\Omega)\int_0^\Omega 
dJ_2P_e(J_2,\Omega)
\delta \left (J-{J_1J_2\over\Omega}\right )\\
\nonumber &+&\left [P_e(\Omega,\Omega)(1-N^2(\Omega))-
P_o(\Omega,\Omega)\right ]P_e(J)\\
&-&\delta(\Omega-J)P_e(\Omega,\Omega),
\end{eqnarray}
\begin{eqnarray}
\nonumber
\lefteqn{-{dP_o(J,\Omega)\over d\Omega}=
P_e(\Omega,\Omega)\int_{-\Omega}^\Omega
dJ_1P_o(J_1,\Omega)}\\
\nonumber &\cdot& \int_{-\Omega}^\Omega dJ_2P_o(J_2,\Omega)
\delta \left (J-{J_1J_2\over\Omega}\right )\\
\nonumber &-&\delta(\Omega-J)P_o(\Omega,\Omega)
+2P_e(\Omega,\Omega)N(\Omega)P_o(-J,\Omega)\\
&+&\left [P_o(\Omega,\Omega)-P_e(\Omega,\Omega)(1-N^2(\Omega))\right ]P_o(J),
\end{eqnarray}
\begin{eqnarray}
\nonumber -{dN(\Omega)\over d\Omega}&=&
\left [P_o(\Omega,\Omega)-P_e(\Omega,\Omega)(1-N^2(\Omega))\right 
]N(\Omega)\\
&+& P_o(-\Omega,\Omega). 
\end{eqnarray}
Here
$P_e(J,\Omega)$ is 
the (normalized) probability distribution of even bonds with $\Omega>J>0$,  
$P_o(J,\Omega)$ is the probability
distribution of odd bonds with $\Omega>J>-\Omega$, and $N(\Omega)$ is the 
fraction of
odd bonds 
that are strongly ferromagnetic $J<-\Omega$. 
$P_o(J,\Omega)$ and $N(\Omega)$ are related by
the normalization condition
$
\int_{-\Omega}^\Omega dJP_o(J,\Omega) + N(\Omega)=1 .
$ 
Anticipating that the bond distribution will become very broad in the low
energy limit, we have neglected factors of order 1 in the strength of 
generated bonds. They become irrelevant in the asymptotic limit\cite{fisher1}.
We have also assumed that the ferromagnetic bonds that are stronger than 
$\Omega$
are {\em much} stronger than $\Omega$, so that two spin-1/2 objects
connected by a strong ferromagnetic bond form a spin-1 object. Again this 
assumption is valid in
the asymptotic limit ($\Omega\rightarrow 0$), 
and simplifies the solution of clusters including 
strong F bonds.

The density, $\rho(\Omega)$, of spins that have not yet  
paired into
singlets at scale $\Omega$ satisfies 
\begin{equation}
-\frac{d\rho(\Omega)}{d\Omega}=-\left\{P_e(\Omega,\Omega)\left[1+N^2(\Omega)
\right ]+
P_o(\Omega,\Omega)\right\}\rho(\Omega). 
\end{equation}
These spins are essentially free at temperatures that are higher 
than $\Omega$.  All thermodynamic quantities can be 
determined from 
$\rho(\Omega)$~\cite{fisher1,thesis,future}.

It is natural to combine the distributions in the following way
\begin{eqnarray}
\nonumber
Q_+(J,\Omega)&=&\frac{1}{1-N(\Omega)}\left 
[P_o(J,\Omega)+P_o(-J,\Omega)\right ],\\
P_-(J,\Omega)&=&P_o(J,\Omega)-P_o(-J,\Omega),
\end{eqnarray}
so that $Q_+(J,\Omega)$ is normalized.  
We find that fixed point solutions to the flow equations are power laws 
$P_e(x)=Pe^{-Px}$ and $Q_+(x)=Qe^{-Qx}$ 
where $x=\log(\Omega/J)$ and $P_-=0$~\cite{thesis,future}.
The variables $P$, $Q$, $N$, and $\rho$ obey
\begin{eqnarray}
\nonumber\frac{dP}{d\Gamma}&=&-N^2P^2 -\frac{(1-N)}{2}QP\\
\nonumber\frac{dQ}{d\Gamma}&=&-(1-N)QP\\
\nonumber\frac{dN}{d\Gamma}&=&(1-N^2)(\frac{Q}{2}-NP)\\
\frac{d\rho}{d\Gamma}&=&-[(1+N^2)P +\frac{1-N}{2}Q]\rho,
\end{eqnarray}
where $\Gamma = \log(\Omega_i/\Omega)$ ($\Omega_i$ is the initial cutoff
of AF bonds).  
We find there are two classes of stable fixed points, corresponding to two 
stable phases. They are the random singlet
phase ($
P=\Gamma^{-1},
Q=Q_0,
N=1,
\rho\propto \Gamma^{-2}$),
and the Haldane phase
($P=P_0,
Q=0,
N=0,
\rho\propto \Omega^{P_0}$, where $0<P_0<1$ is a nonuniversal number).
There is also an unstable fixed point (near which there is
a single relevant operator):
$(P=Q=2\Gamma^{-1},
N=\frac{1}{2},
\rho\propto \Gamma^{-3})$. It describes the critical point.
In the following we describes the physical nature of these fixed points.

In the random singlet phase, all the odd bonds become F bonds much stronger
than the AF even bonds. All the spin-1/2's are ferromagnetically 
combined into spin-1's.  The spin-1's then couple into singlets 
over all
length scales.  
The ground state and thermodynamic properties are the same as
for the spin-1/2 random singlet state studied
previously~\cite{mdh,fisher1}.  The disorder averaged spin-spin correlation 
function
$C(r)$ decays as $r^{-2}$, and the susceptibility take the universal form
in the low temperature limit: $\chi\sim [T\log^2T]^{-1}$.   
There is no gap because the bond
distribution has weight at $J=0$. 

In the Haldane phase all odd bonds (F and AF) become {\em much weaker}
than the even bonds, only spin-1/2's
remain in the system and
they form singlets only over even bonds. 
The system may be viewed 
as a set of uncoupled dimers.  
The spin-spin correlations decay exponentially with a
finite correlation length. There is also long range string-topological 
order\cite{hyman}.
This phase is analogous to the random
dimer phase in random dimerized
AF
spin-1/2 chains\cite{hyman} and the ground state resembles the valence bond
solid state.
The flow equations describe the Griffiths region of the Haldane phase where
there is no gap and the susceptibility diverges as a power law with a
nonuniversal exponent $\chi\sim T^{-(1-P_0)}$.
The flow equations are
only valid when the disorder is broad, and so can not describe the
crossover from gapped to gapless behavior within the Haldane phase as
the randomness is increased.

To determine critical exponents we consider small perturbations near
the unstable fixed point:
\begin{eqnarray}
\nonumber P=\frac{2}{\Gamma}(1 + \delta_p\Gamma^{\lambda})\\
\nonumber Q=\frac{2}{\Gamma}(1 + \delta_q\Gamma^{\lambda})\\
N=\frac{1}{2}(1 + \delta_n\Gamma^{\lambda})
\end{eqnarray}
and
expand the flow equations
to linear order in $\delta$. 
There are two irrelevant perturbations 
($\lambda=-1,\lambda=\frac{-1 - \sqrt{13}}{2}$) and one 
relevant
perturbation ($\lambda_{+}=\frac{-1 + \sqrt{13}}{2}$).  
For relevant flows,
if $\delta_n>0$, the odd bonds are stronger than the even bonds, the
density of spin-1's is increasing, and the system flows 
to the random singlet fixed point.  If $\delta_n<0$ the even bonds are
stronger than the odd bonds, the density of spin-1's is decreasing, and
the system flows to the random Haldane phase.
The crossover from critical to Haldane behavior occurs at the energy scale 
where 
$\delta\Gamma^{\lambda}\approx 1$.  The energy scale at which this occurs
is
$\Gamma_0= \delta^{-\frac{1}{\lambda_+}}$.  The density of spins at this scale 
is $\rho_0=\Gamma_0^{-3}$, so the correlation length is
$\xi\approx\rho_0^{-1}=\delta^{-\nu}$ where 
$\nu= \frac{3}{\lambda_+}\approx 2.3$. 
Following the same analysis as in Ref.~\cite{hyman}, the
string-topological order parameter defined as~\cite{rn}
\begin{equation}
{{T}}=\lim_{j-i\rightarrow\infty}\left\langle\Psi_0\left|S^z_i
\exp\left[i\pi\sum_{i<k<j}{S^z_k}\right]
S^z_j\right|\Psi_0\right\rangle,
\end{equation}
scales as
$
{{T}}\propto (-\delta_n)^{2\nu},
$
in the Haldane phase near the critical point.
Just like the RS fixed point for the spin-1/2 chain\cite{fisher1}, the 
bond distributions become infinitely broad on a logarithmic scale. Thus
our scheme becomes asymptotically exact at the critical point, and the 
critical exponents we find here are exact.

Before closing we briefly discuss relations between the present work and some
existing papers. Our model is related, but different from that studied
by Westerberg {\em et al.}\cite{westerberg}.
In their model, ferromagnetic bonds may appear
in both even and odd bonds, and spins of arbitrarily large size appear at
low energy. The special topological structure of our model prevents this
from happening in the present system.
Boechat {\em et al.}\cite{boechat} also anticipated the existence of
a random singlet phase for strong randomness. 
As in our previous work\cite{kunspaper}, they also find {\em spontaneously}
dimerized chains are unstable against {\em weak} randomness.
They did not, however, address
the weak randomness regime of the spin-1 chain, and the phase transition that
we discuss here.
The correlation length exponent $\nu\approx 2.3$ we find here is extremely
close to that of the delocalization transition in integer quantum Hall systems.
Lee\cite{lee} showed that the integer quantum Hall transition may be
mapped onto the dimerization transition in the pure SU(0) spin chain.
It is unclear at this stage, whether this is merely a coincidence, or there
exists a fundamental physical reason, that these two apparently different 
transitions have the same critical exponent.

To summarize, 
in this letter we determined the critical properties of the randomness driven 
phase transition of the spin-1 chain.  For weak randomness, the spin-1 chain is 
in the Haldane phase and the ground state resembles the valence bond solid 
state.  The ground state has topological order, spin-spin correlations 
decays exponentially and there is a gap to the 
excited states.  For broader distributions the gap is filled in and the 
correlation length increases but the topological order persists.  Eventually 
the 
critical point is reached and the 
correlation length diverges. 
The correlation length exponent is $\nu=\frac{6}{\sqrt{13} -1} \approx 2.3$.
Beyond the critical point the topological order 
is 
zero, the disorder averaged spin-spin 
correlations decay algebraically, and the ground state resembles the random 
singlet state.

R.A.H. and K.Y. were supported, respectively,  
by National Science
Foundation Grants DMR-9224077 and DMR-9400362. 
The authors thank S. M. Girvin, R. N. Bhatt, and
D. S. Fisher for valuable discussions.

\end{document}